  \documentstyle[aps,multicol]{revtex}
  \input{epsf}
\begin{document}
\draft
\title{Magnetotransport in manganites and the role of quantal phases II: Experiment}
\author{S. H. Chun, M. B. Salamon, P. D. Han, Y. Lyanda-Geller, and P. M. Goldbart}
\address{Department of Physics and Materials Research Laboratory, University of\\
Illinois at Urbana-Champaign, Urbana, Illinois 61801-3080}
\date{April 22, 1999}
\maketitle

\begin{abstract}
As in conventional ferromagnets, the Hall resistivity $\rho _{xy}$ of a La$%
_{2/3}$(Ca,Pb)$_{1/3}$MnO$_3$ single crystal exhibits both ordinary and
anomalous contributions at low temperature. However, these contributions,
unexpectedly, have opposite signs. Near $T_c$, the ordinary contribution is
no longer evident and $\rho _{xy}$ is solely determined by the sample
magnetization, reaching an extremum at $\approx $40 \% of the saturated
magnetization. A new model for the anomalous Hall effect, incorporating the
quantal phase accumulated by double-exchange, three-site hopping reproduces
this result. Below $T_c$, $\rho _{xy}$ reflects the competition between
normal and anomalous Hall effects.
\end{abstract}


\begin{multicols}{2}
\narrowtext
Among the many intriguing properties exhibited by doped perovskite
manganites, perhaps none is more puzzling than the Hall effect. A number of
measurements have been reported on various members of the series La$_{1-x}$A$%
_x$MnO$_3$ (where A is Ca, Sr, or Pb), and all show common anomalous features%
\cite{Snyder,Jaime,Matl,Jacob,Asamitsu,ChunLPMO}. At the lowest
temperatures, the Hall resistivity $\rho _{xy}$ is positive and linear in
field, as expected for hole-doped materials, but rather smaller than
expected. \ This has recently been attributed to charge compensation and
Fermi-surface shape effects\cite{ChunLPMO}. As the temperature is increased,
a component of the Hall resistivity appears that is proportional to the
magnetization $M(H,T)$, but has a negative sign. The appearance of an
anomalous Hall effect is, of course, commonly observed in ferromagnets, but
is usually attributed to spin-orbit scattering of the charge carriers and
normally carries the same sign as the ordinary Hall contribution. A strong
negative contribution to $\rho _{xy}$ persists through the transition
temperature, but loses its proportionality to $M,$ until, at temperatures $%
\geq 1.5T_c,$ $\rho _{xy}$ becomes linear in field again (though negative)
with a slope decreasing exponentially with increasing temperature in the
manner expected for the Hall resistivity of small polarons\cite
{Jaime,ChunLCMO,Holstein}. Experimental data \cite{ChunLPMO} suggest that
charge transport is not polaronic in the temperature range $T_c-1.5T_c$.

In this Letter, we present new Hall resistivity data on optimally doped
manganite single crystals, with emphasis on the temperature region between
the band-like, positive Hall regime at low temperatures and polaronic
behavior at high temperatures. We show that the Hall resistivity $\rho _{xy}$
is a function only of $M(H,T),$ reaching an extremum near $M/M_{\text{sat}%
}=0.4$ when this value can be reached with laboratory fields at temperatures
above the Curie temperature $T_c.$ In fact, as we shall see, the data for
all temperatures $T\geq T_c$ lie on a universal curve that follows from the
theoretical model presented in a companion paper that we refer to as I \cite
{Yuli}. Data taken below $T_c$ track this universal curve once the
magnetization is saturated, but are shifted to slightly larger values of $%
\left| \rho _{xy}\right| .$ We argue that this is due to the return of band
conduction as ferromagnetism, driven by double exchange, sets in.

  \begin{figure}[hbt]
  \epsfxsize=7.5truecm
 \centerline{\epsfbox{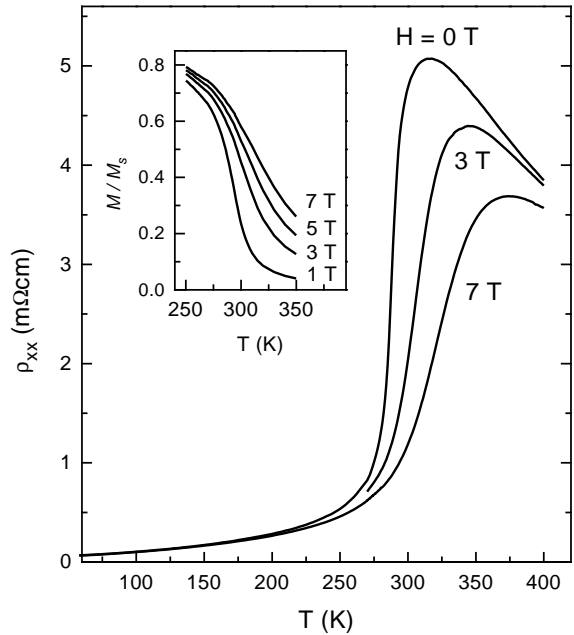}} 
  \vskip+0.20truecm
  \caption{Main panel: the temperature dependence of longitudinal 
resistivity $\rho_{xx}(H,T)$ of a ${\rm La_{2/3}(Pb,Ca)_{1/3}MnO_3}$ 
single crystal for different fields. Inset: the temperature dependence 
of magnetization $M(H,T)$
for the same crystal.}
  \label{fig1}
  \end{figure}

High quality single crystals of La$_{2/3}$(Ca,Pb)$_{1/3}$MnO$_3$ were grown
from 50/50 PbF$_2$/PbO flux. It was found that the addition of Ca favors
optimally doped crystals; chemical analyses of crystals from the same batch
gave the actual composition as La$_{0.66}$(Ca$_{0.33}$Pb$_{0.67}$)$_{0.34}$%
MnO$_3.$ Specimens for the Hall measurements were cut along crystalline axes
from larger, pre-oriented crystals. Details of the measurement technique and
analysis of the low temperature region have been presented elsewhere\cite
{ChunLPMO}. The Hall resistivity $\rho _{xy}$ and longitudinal resistivity $%
\rho _{xx}$ were measured simultaneously as functions of field and
temperature. The magnetization of the same sample was measured following the
Hall experiment, and was used to correct for demagnetization fields. Figure
1 shows the longitudinal resistivity as a function of temperature at zero
field, 3 T and 7 T. Magnetization curves are shown in the inset. The
residual resistivity of this sample, $\rho _{xx}^0\approx 51$ $\mu \Omega $
cm, is comparable to the best values obtainable in these materials,
indicating the absence of grain boundaries in our sample. The maximum change
in resistivity with temperature $d\rho _{xx}/dT$ occurs at 287.5 K in zero
field, moving to higher temperatures with increasing field. This gives rise
to the ``colossal magnetoresistance (CMR)'' effect, which is 326\% at 293 K
and 7\ T. A scaling analysis of the magnetization data very close to the
metal-insulator transition (MIT) gives a Curie temperature of $T_c=285$ K,
but this must be taken cautiously as the scaling exponents differ
significantly from those expected from a 3D Heisenberg ferromagnet.
Nevertheless, it is clear that $\rho _{xx}$ and $M$ are closely correlated
in this system.

  \begin{figure}[hbt]
  \epsfxsize=7.5truecm
 \centerline{\epsfbox{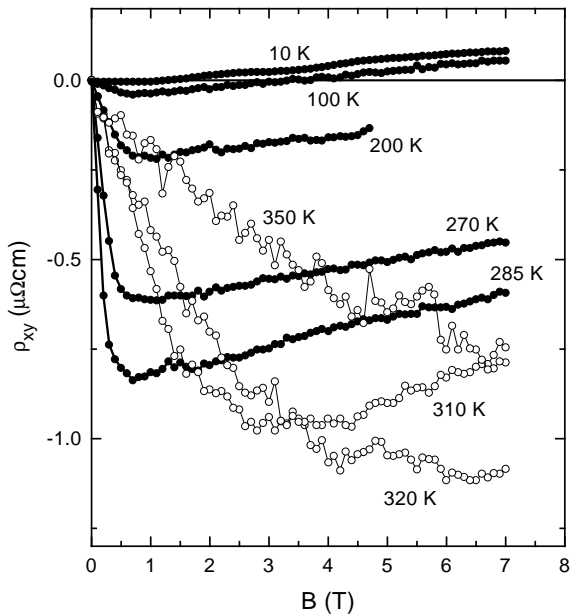}} 
  \vskip+0.20truecm
  \caption{Hall resistivity $\rho_{xy}$ of the same crystal as a 
  function of field at indicated temperatures.}
  \label{fig2}
  \end{figure}

In Fig.~\ref{fig2}, 
we show the field dependence of $\rho _{xy}$ at a number of
temperatures.{\em \ }As noted above, the Hall resistivity is positive and
linear in field at low temperatures, indicating that the anomalous Hall
effect (AHE) is small. As the temperature is increased, the AHE contribution
increasingly dominates the ordinary Hall effect (OHE), first causing $\rho
_{xy}$ to change sign as a function of field, and then driving it negative
for all fields in the range measured. In ferromagnetic metals, the Hall
resistivity is generally written as\cite{Hurd} 
\begin{equation}
\rho _{xy}=R_H[\mu _0H_{app}+\mu _0(1-N)M]+R_S\mu _0M,  \label{eq1}
\end{equation}
where $R_H$ is the coefficient of the OHE, while $R_S$ is the coefficient of
the anomalous contribution. In Fig. 2, we have plotted the data in terms of
the internal field, given by the square brackets in Eq. (\ref{eq1}), using a
demagnetization factor $N$ calculated from the dimensions of the sample. As
the temperature is increased through the transition temperature, the minimum
of $\rho _{xy}$ moves to higher fields, and the positive high-field
contribution disappears. That there are rapid changes near $T_c$ is not
surprising because, even in an ordinary ferromagnet, $R_S$ depends on
longitudinal resistance which, in these samples, changes dramatically with
temperature and applied field. However, $R_S$ is then attributed to
scattering from spin disorder, via either a skew-scattering \cite{Maranzana}
or side-jump processes \cite{Berger}, both of which require that the
resistance be dominated by spin-disorder scattering. Even at low
temperatures, where $R_S$ is proportional to $\rho _{xx},$ the sign is
opposite that expected from skew-scattering theory \cite{Maranzana}.
Further, as has been pointed out by many authors, the resistance changes
observed here are too large to be the result of spin-dependent scattering,
and must involve some form of localization, an effect we will invoke to
explain the changes apparent here.

As discussed in I, we assume that transport in the transition region is
dominated by hopping processes, giving rise to a longitudinal conductivity $%
\sigma _{xx}=(ne^2d^2/k_BT)W_0\cos ^2(\theta /2),$ where $d$ is the distance
between ions. Here $W_0$ is the probability of phonon-assisted direct hops
and we have explicitly separated the Anderson-Hasegawa factors $\cos
^2(\theta /2)$. The AH conductivity, correspondingly, is given by $\sigma
_{xy}=(ne^2d^2/k_BT)W_1$, where $W_1$ is the probability of hopping between
two ions via an intermediate state on a third ion and includes
Anderson-Hasegawa factors [see Eq.~(1) in I]. The problem then reduces to
determining the ratio between direct and indirect hopping rates as a
function of the spin texture. Because $W_1$ involves two-phonon processes,
we write $W_1/W_0^2=\alpha \hbar \zeta /k_{{\rm B}}T$, where $\alpha $ is a
numerical factor describing the multiplicity of the various carrier-phonon
interference processes (see~\cite{Holstein}), the number of intermediate
sites, and the difference between nearest- and next-nearest-neighbor hopping
amplitudes, and $\zeta $ is an asymmetry parameter. For the OHE, $\zeta
\propto \sin ({\bf B}\cdot {\bf Q}/\phi _0)$, where ${\bf Q}$ is the area
vector of the triangle enclosed by the three sites. In the AHE case, it
follows from Eqs.(2) and (4) in~I that $\zeta \simeq 3[{\bf g}_{jk}\cdot (%
{\bf n}_j\times {\bf n}_k)][{\bf n}_1\cdot ({\bf n}_2\times {\bf n}_3)]/4$,
where ${\bf g}_{jk}$ are characteristic vectors arising from the spin-orbit
quantal phase in the hopping amplitude; ${\bf n}_j$ are unit vectors of the
core spins in the triad, and ${\bf n}_1\cdot ({\bf n}_2\times {\bf n}_3)$ is
the volume of a parallelepiped defined by core-spin vectors, denoted as $q_P$
in I. The anomalous Hall resistivity can be written in the simple form 
\begin{equation}
\rho _{xy}\simeq -\sigma _{xy}/\sigma _{xx}^2=-\frac 1{ne}\left( \frac{%
\alpha \hbar \zeta }{ed^2}\frac 1{\cos ^4(\theta /2)}\right) .  \label{eq2}
\end{equation}

The evaluation of Eq.(\ref{eq2}) reduces to a determination of $\cos (\theta
/2)$ and products $({\bf n}_j\times {\bf n}_k)$ and ${\bf n}_1\cdot ({\bf n}%
_2\times {\bf n}_3)$ that survive averaging over all possible triads. In
contrast to the hopping OHE in doped semiconductors \cite{Galperin}, where
only two sites in an optimal OHE triad are connected to the conducting
network (CN), all three triad sites must participate in the network if they
are to contribute to the AHE. Our argument is that if one of the sites is
not a part of the CN then its core spin must be roughly opposite that of the
other two spins, yielding a vanishingly small $q_P$. It is reasonable then
to assume that the CN is formed by ions with splayed core spins oriented
roughly in the direction of average magnetization ${\bf m}$. We then
consider the square lattice formed by Mn ions in planes perpendicular to $%
{\bf m}$, and assume that the core spin vectors of the four ions in a
typical elementary plaquette belonging to CN lie equally spaced on the cone
whose half angle is given by $\beta =$ $\cos ^{-1}[M(H,T)/M_{\text{sat}}].$
A typical pair of ions that determine the longitudinal current, and a
typical triad can now be chosen from ions of this plaquette. From elementary
geometry, it follows that $2\cos ^2({\theta /2)}=1+\cos ^2\beta $, $%
q_P=2\cos \beta \,\sin ^2\beta $, and ${\bf m\cdot }({\bf n}_j\times {\bf n}%
_k)=\sin ^2\beta $. To discuss the AHE magnitude, we need first to estimate
the characteristic values of $|{\bf g}_{jk}|\sim g$ arising from the
spin-orbit interaction (SOI). As we discussed in~I, the SOI term leads to a
Dzyaloshinski-Moriya contribution to the eigenenergy of carriers, whose
magnitude is given by $g\sim Ze^2/4m_ec^2d_0$, where $d_0$ is the radius of
an Mn core d-state. An estimate based on free electron parameters gives $%
g\sim 5\times 10^{-4}$. While renormalization of carrier parameters in
crystals may tend to increase $|{\bf g}_{jk}|,$ it is necessary to allow
admixtures of core orbitals with outer-shell wavefunctions in order to have $%
|{\bf g}_{jk}|\neq 0$ for symmetric potentials. The non-collinearity of the
Mn-O-Mn bonds that allows carrier hopping around triads\ (including jumps
along plaquette diagonals) effectively generates such an admixture. Thus, a
value of $g\sim 5\times 10^{-4}$ is reasonable.

  \begin{figure}[hbt]
  \epsfxsize=7.5truecm
 \centerline{\epsfbox{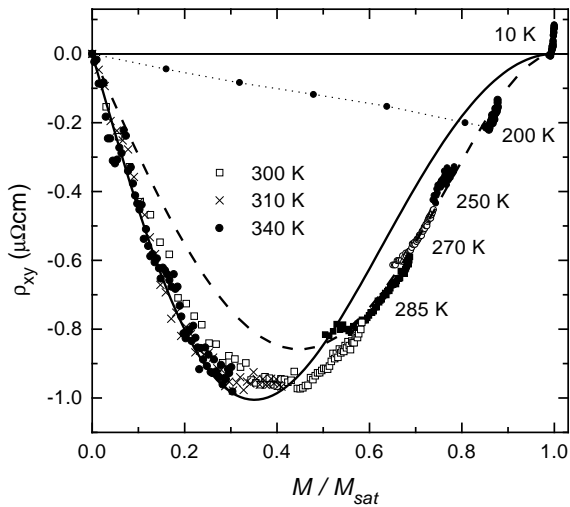}} 
  \vskip+0.20truecm
  \caption{Scaling behavior between $\rho_{xy}$ and sample magnetization $M$.
   The solid line is a fit to Eq. (3); the dashed line is the numerator of 
   Eq.~(3) only. There are no fitting parameters except normalization.}
  \label{fig3}
  \end{figure}

As discussed in I, the magnitude of the longitudinal (and anomalous Hall)
resistivities in the regime of abrupt increase of the resistivity depends
not only on properties of individual pairs (triads), but also on how are
they connected to the CN. We estimate the macroscopic longitudinal and Hall
conductivities at the low temperature limit of our model, where the CN is
still fully connected. Taking $n=5.6\times 10^{21}$ cm$^{-3}$, $W_0\sim
2.5\times 10^{13}$ s$^{-1}$, and $\cos \beta =0.6$ from the magnetization
data at $T=275$ K (Fig. 1), we obtain $\rho _{xx}\simeq 1$ m$\Omega $ cm
which coincides with the value of the experimentally observed resistivity
(Fig. 1). The AHE contribution to the Hall resistivity, assuming numerical
factor $\alpha =2.5$, is then $\rho _{xy}=-0.5$ $\mu \Omega $ cm, in
agreement with the experimentally observed Hall resistivity at the same $T$
(Fig. 2). The equivalent expression for the hopping Hall resistance in the
Holstein mechanism has $\zeta \simeq \cos ^2(\theta /2)\cos \beta \sin ({\bf %
B}\cdot {\bf Q}/\phi _0)$ and, at $B=1$ T, is an order of magnitude smaller
than the AHE. We expect the macroscopic, hopping AHE and OHE to have the
same sign, opposite that of the OHE\ in the metallic regime.

To relate $\rho _{xy}$ to $m\equiv \left| {\bf m}\right| $, we introduce a
percolation factor $P$ for $\sigma _{xx}$ describing the connectivity of the
pair to the CN; for the AH conductivity the corresponding factor would be $%
P^2$ because both pairs in a triad must, as discussed above, belong to the
CN. It is remarkable that throughout the localization regime, $\rho _{xy}$
is, nevertheless, determined by currents formed in individual pairs and
triads, because the factors of $P$ cancel. Therefore, as long as $q_P$ and
the angles between neighboring spins can be directly related to $m\equiv
M/M_{\text{sat}}=\cos \beta $, $\rho _{xy}$ depends on $H$ and $T$ only
through $m(H,T),$ and is given by. 

\begin{equation}
\rho _{xy}=\rho _{xy}^0\frac{m(1-m^2)^2}{(1+m^2)^2}  \label{eq3}
\end{equation}

The corresponding curve is shown in Fig.~3, where the data of Fig.~2 are
replotted as a function of ${M/M_{\text{sat}}}$. At and above $T_c$ the data
fall on a smooth curve that reaches an extremum at $M/M_{\text{sat}}\simeq
0.4.$ Below $T_c$ the data first change rapidly with magnetization as
domains are swept from the sample before saturating and following the
general trend. At the lowest temperatures, the metallic OHE appears as a
positive contribution at constant magnetization. The solid curve in Fig. 3
follows Eq.~(\ref{eq3}) with $\rho _{xy}^0=-4.7$ $\mu \Omega $ cm,
consistent with the estimates of $\rho _{xx}$ and $\rho _{xy}$ given above.
Down to 285 K, which is the Curie temperature determined by scaling
analysis, Eq. (\ref{eq2}) describes the data reasonably well. In addition,
the extremum is located at $M/M_{\text{sat}}=\cos \beta \approx 0.35$, close
to the experimental extremum. Below $T_c,$ the longitudinal resistivity is
metallic and no longer dominated by magnetic disorder. However, local spin
arrangements still dominate the AHE via asymmetric scattering, as discussed
in a forthcoming paper. The numerator of Eq.(\ref{eq3}), $m(1-m^2)^2,$
essentially the behavior of $\sigma _{xy}$ alone, has an extremum at $m=1/%
\sqrt{5}\simeq 0.45$ as shown by the dashed line in Fig. 3. 
The broader maximum in
the data suggest a shift toward a hopping model for $\rho _{xx}$ and $\rho
_{xy}$ as the sample is warmed through the metal-insulator transition.

To consider the effect of OH contributions, which are masked in $\rho _{xy}$
by the large magnetoresistance, it is useful to examine the field and
temperature dependence of $\sigma _{xy}$ instead. As seen in the main panel
of Fig. 4, the magnetic field dependence of $\sigma _{xy}$ at 200 K clearly
shows the OHE by free carriers at high fields, opposite in sign to the AHE.
The AHE sign can be inferred by extrapolating the high field curve to its $%
B=0$ intercept. At $T=200$ K, $\rho _{xx}$ is mainly metallic. At 265 K,
where $\rho _{xx}$ starts to increase rapidly, $\sigma _{xy}$ saturates at
external magnetic fields $H\sim 2$ T. At that magnetic field $M\sim 0.7M_{%
\text{sat}}$, relatively close to maximal value achievable at this
temperature, $M\sim 0.8M_{\text{sat}}$ at 7 T. This saturation effect
strongly suggests that the AHE is dominant and that the negative Holstein
OHE \cite{Holstein} is either suppressed or partly compensated by the
decrease in the AHE from reductions in $q_P$ by the magnetic field. As an
increase in the magnetic field tends to delocalize carriers, we may expect
to see the onset of metallic OHE at larger applied fields.

  \begin{figure}[hbt]
  \epsfxsize=7.5truecm
 \centerline{\epsfbox{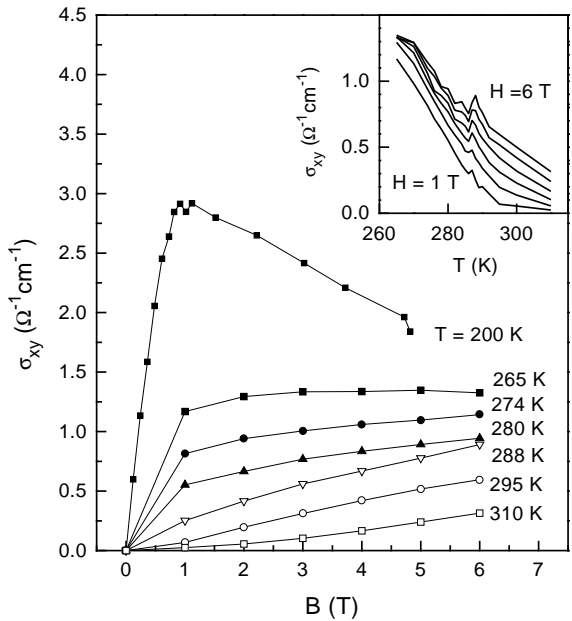}} 
  \vskip+0.20truecm
  \caption{Hall conductivity $\sigma _{xy}(H,T)=-\rho _{xy}/\rho _{xx}^2$ 
  of the same crystal. Main panel shows the field dependence. The inset 
  shows the temperature dependence for $H=$ 1, 2, 3, 4, 5, 6 T from bottom 
  to top.}
  \label{fig4}
  \end{figure}

Another interesting feature is that the temperature dependence of $\sigma
_{xy}$ shows an anomaly at the same temperature as the zero field $d\rho
_{xx}/dT$ peak, (Fig. 4, inset). The size of the anomaly increases with
increasing field, implicating the OHE. If this is related to a polaronic
collapse of the conduction band\cite{Millis}, the anomaly should shift as $H$
increases. However, peak shifts much less than does the $d\rho _{xx}/dT$
peak (see Fig. 1). To the extent that the transition is a percolation
process in which metallic regions grow to form a percolation network, it is
possible for the non-metallic Hall contribution to remain dominant down to
the percolation threshold\cite{bergman}, resulting in the sudden appearance
of the OHE when the system becomes fully metallic. Indeed, this effect is
evident, though less dramatic, in the Hall resistivity of Fig. 3, as a
deviation of the data below $T_c$ from the universal curve deduced from the
quantal phase calculation.

In conclusion, we measured the Hall resistivity, the longitudinal
resistivity, and the magnetization of a La$_{2/3}$(Ca,Pb)$_{1/3}$MnO$_3$
single crystal. Very similar results have been observed in single crystals
of Ca- and Sr-doped LaMnO$_3$ and will be reported elsewhere. We find that
the Hall resistivity is solely determined by the sample magnetization ($M$)
near and somewhat above the transition temperature. A model for the AHE,
based on the Holstein picture in which interference between direct hops and
those via a third site provides the necessary quantal phase\cite{Yuli},
explains the results quite well. Unlike Holstein polarons, an additional
phase here is introduced by the strong Hund's rule coupling that forces the
hopping charge carrier to follow the local spin texture. Below the
transition temperature, the AHE competes with the OHE as long-range magnetic
order and, presumably, an infinite percolating metallic cluster, develops. A
sharp, field dependent drop in the Hall conductivity at the transition
temperature is qualitative evidence for this cross over between
hopping-dominated Hall effects and features similar to those observed in
more conventional ferromagnets.

This work was supported in part by DOE DEFG-91ER45439.

\end{multicols}
\end{document}